\documentstyle[epsf]{aipproc} 
\def\psim{\lower.5ex\hbox{$\; \buildrel \propto \over\sim \;$}} 
\def\gtrsim{\lower.5ex\hbox{$\; \buildrel > \over\sim \;$}} 
\def\lesssim{\lower.5ex\hbox{$\; \buildrel < \over\sim \;$}} 
\def\e{{\epsilon}} 
\def\ep{{\epsilon^\prime}} 
\def\gg{{\gamma\gamma}} 
 
\begin{document} 
\title{High Energy Radiation from \\Gamma Ray Bursts} 
 
\author{Charles D. Dermer$^*$\thanks{Work supported by the Office of Naval Research.} and James Chiang$^{\dagger}$} 
\address{$^*$Naval Research Laboratory, Code 7653, Washington, DC  20375-5352\\ 
$^{\dagger}$JILA, University of Colorado, Campus Box 440, Boulder, CO 80309-0444} 
 
\maketitle 
 
\begin{abstract} 
Gamma-ray burst (GRB) engines are probed most intimately during the prompt gamma-ray luminous phase when the expanding blast wave is closest to the explosion center. Using GRBs 990123 and 940217 as guides, we briefly review observations of high-energy emission from GRBs and summarize some problems in GRB physics. $\gg$ transparency arguments imply relativistic beaming.  The parameters that go into the external shock model are stated, and we show numerical simulation results of gamma-ray light curves from relativistic blast waves with different amounts of baryon loading. A distinct component due to the synchrotron self-Compton process produces significant emission at GeV and TeV energies.  Predictions for spectral and temporal evolution at these energies are presented for a blast wave expanding into uniform surroundings.  Observations of the slow decay of GeV-TeV radiation provide evidence for ultra-high energy cosmic ray acceleration in GRBs. 
 
\end{abstract} 
 
\section{Introduction} 
 
The cosmological hypothesis for the origin of GRBs has been favored ever since BATSE showed that GRB sources were isotropically distributed about us, yet were bounded in spatial extent \cite{meegan92}. Burst studies were revolutionized by the discovery of X-ray afterglows \cite{cea97} with the Beppo-SAX satellite.  The good X-ray imaging of the Narrow Field Instrument on Beppo-SAX quickly led to the identification of optical \cite{vea97} and radio counterparts \cite{fea98}, permitting redshift measurements from optical transient counterparts and directionally coincident host galaxies. Once the redshift is known, the power and energy release follow modulo the collimation factor $\delta \Omega/4\pi$.  
  
\begin{table} 
\caption{Inferred Isotropic 50-300 keV$\times(1+z)$ Luminosities and Energy Releases from a Sample of GRBs with Redshifts} 
\label{table1} 
\begin{tabular}{lddddd} 
   GRB& ~~Redshift $z$~~ &  
   \multicolumn{1}{c}{Peak Flux\tablenote{Units of photons cm$^{-2}$ s$^{-1}$ in 50-300 keV range.}} ~~ & 
\multicolumn{1}{c}{Fluence\tablenote{Units of $10^{-6}$ ergs cm$^{-2}$ in 50-300 keV range, except for GRB 970228 (35-1000 keV) and GRB 971214 ($> 20$ keV).}} ~~& \multicolumn{1}{c}{$L_\gamma$\tablenote{Assuming an $\Omega_m = 0.3$, $\Omega_\Lambda = 0.7 $ cosmology with $H_0 = 65$ km s$^{-1}$ Mpc$^{-1}$, and a mean photon energy of 107 keV implied by a flat $\nu F_\nu$ spectrum.} ($10^{51}$ ergs s$^{-1}$)}~~& 
  \multicolumn{1}{c}{$E_\gamma$ ($10^{52}$ ergs)}\\ 
\tableline 
970228 & 0.695 & 3.5 & 6.1 & 1.1 & 0.65\\ 
970508 & 0.835 & 1.2 & 3.1 & 0.57 & 0.47\\ 
971214 & 3.418 & 1.95 & 10.9 & 24 & 17\\ 
980425 & 0.0085 & 0.96 & 4.0 & 6.2$\times 10^{-5}$ & 7.3$\times 10^{-5}$ \\ 
980703 & 0.966 & 2.4 & 45.9 & 1.6 & 9.1\\ 
990123 & 1.60 & 16.4 & 509 & 35 & 240\\ 
\end{tabular} 
\end{table} 
 
Table 1 lists soft gamma-ray luminosities and energies for a sample of the dozen GRBs now known with measured redshifts. In the case of GRB 990123, the directional $\gamma$-ray power and energy reach values as large as $\partial L_\gamma /\partial \Omega \sim 3\times 10^{51}$ ergs s$^{-1}$ sr$^{-1}$ and $\partial E/\partial \Omega \sim 2\times 10^{53}$ ergs sr$^{-1}$\cite{kea99}, respectively.  Fig. 1 shows the 50-300 keV light curve (left) and $\nu F_\nu$ spectrum\cite{bea99} (right) of GRB 990123. The spectrum is summed over the 12.3 - 45.1 s interval after the trigger time\footnote{Except for EGRET's Total Absorption Shower Counter (TASC) spectrum, which accumulated photons from -0.057 s to 64.5 s.}. Given this luminosity and the intrinsic variability time scale $t_v \sim 5/(1+z)$ s, the compactness parameter $\ell = L\sigma_T/(4\pi m_ec^3 ct_v) \sim 10^{12}$ is enormous, so that gamma rays could not escape without invoking directed beams of photons and directed relativistic motions of the emitting particles.
\begin{figure}[b!] 
\vskip-0.4in
\centerline{
\epsfxsize=0.4\textwidth\epsfbox{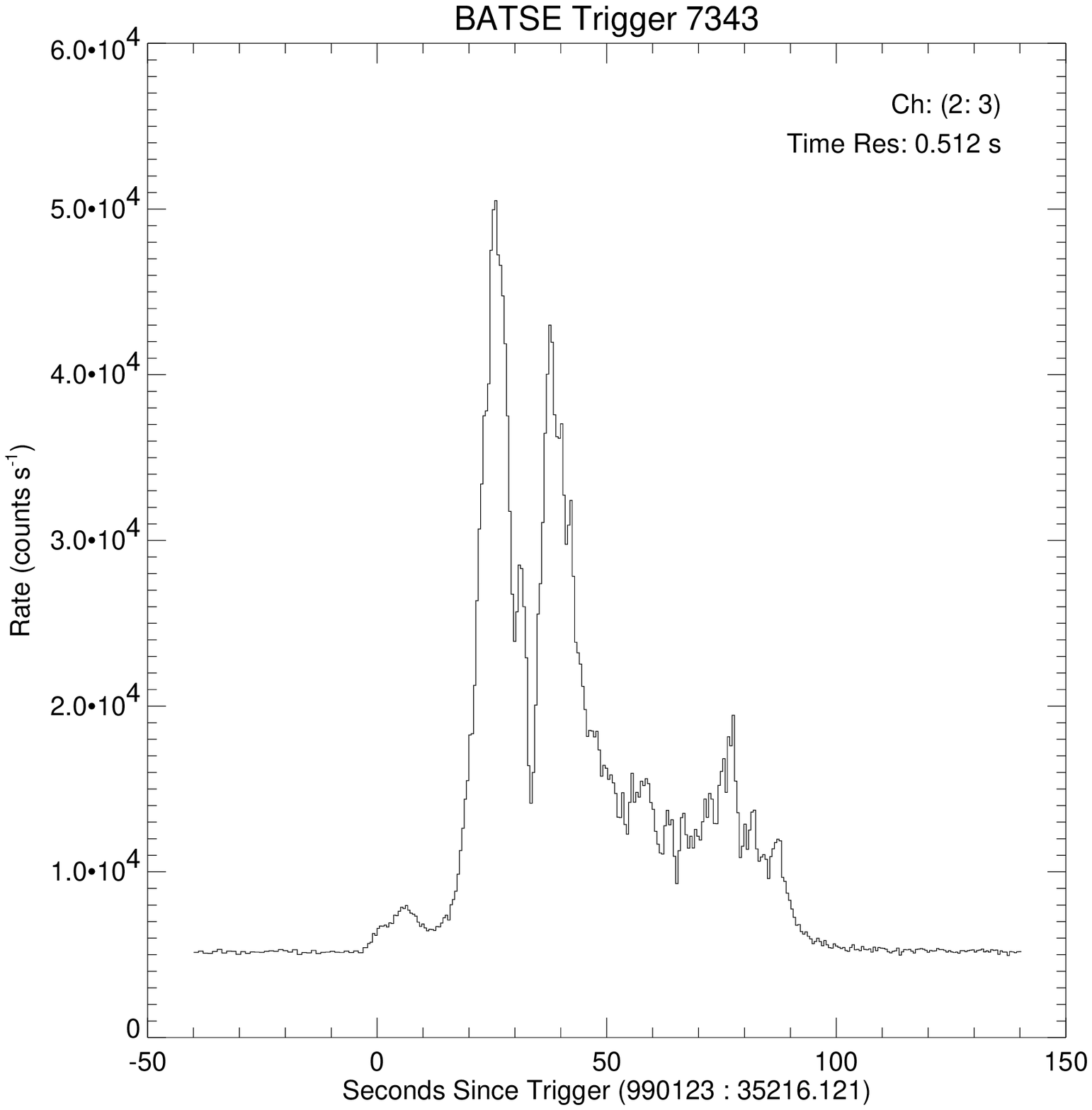}
\epsfxsize=0.36\textwidth\epsfbox{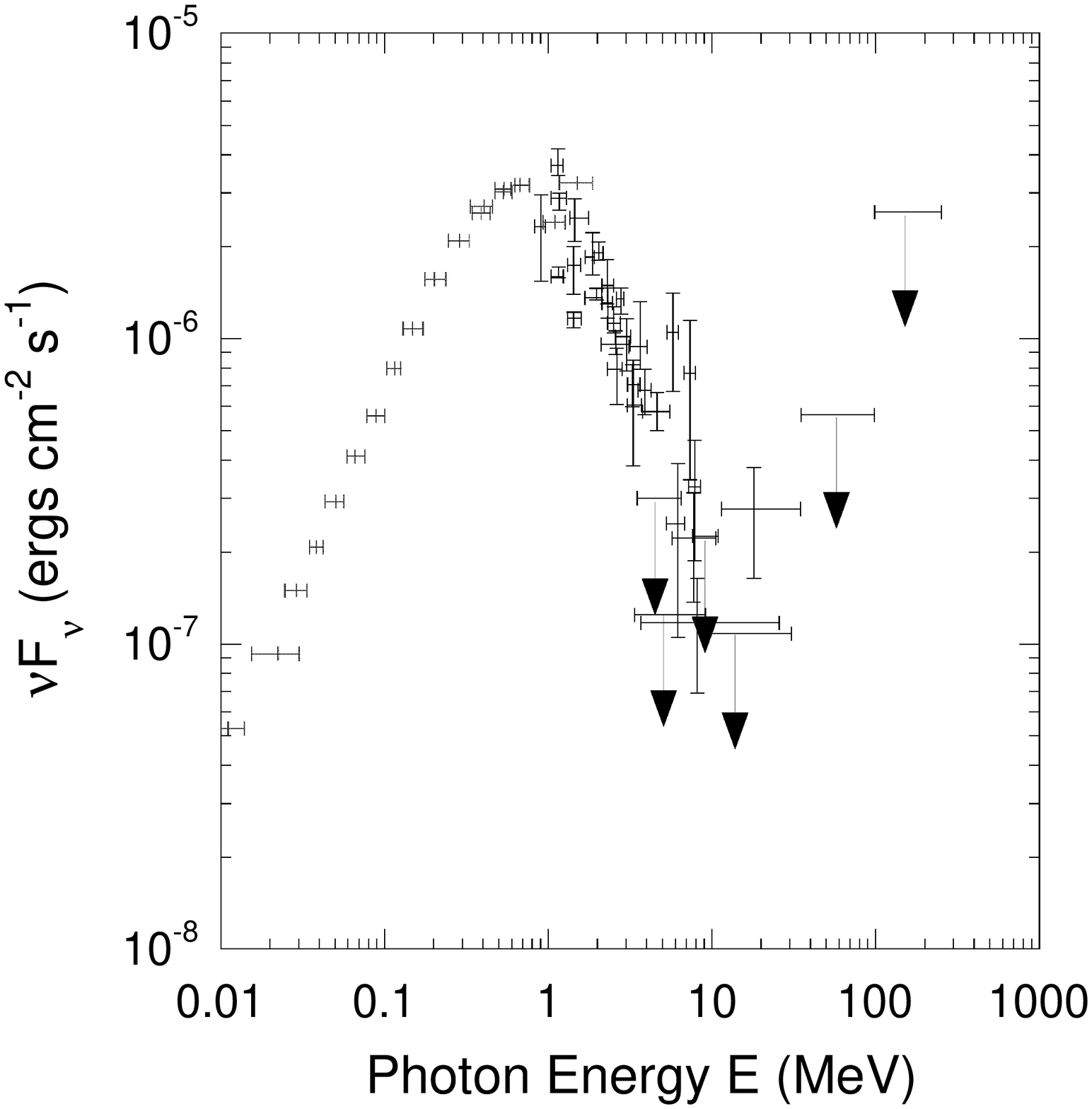}
}
\caption{(left) The gamma-ray light curve of GRB 990123, observed with BATSE in the energy range 50-300 keV. (right) The time-averaged broadband spectrum of GRB 990123.}
\label{fig1}
\end{figure}
 
The inference of bulk relativistic motion from the gamma-ray observations anticipated the expanding relativistic blast wave model that so readily explains temporal X-ray and optical power-law afterglow decays \cite{v97,w97,wmr97}.  In the external shock model (ESM), a relativistic blast wave is energized as it passes through and captures material from the surrounding gas and dust \cite{rm92,mr93}. In the colliding shell (internal shock) model \cite{kps97,dm98}, the engine's activity is prolonged and intermittent. The ESM explains the long wavelength afterglow behavior and, as argued elsewhere \cite{dbc99,bd00}, the phenomenology of the prompt gamma-ray luminous phase and short timescale variability\cite{dm99}.    
 
Here we describe the potential of gamma-ray observations to characterize properties of GRB sources.  In Section 2, a brief summary of GRB gamma-ray observations is given. Some unsolved GRB problems are mentioned in Section 3.  Section 4 spotlights the method of inferring properties of the blast wave from $\gg$ transparency arguments. This leads to a description of the standard fireball/blast-wave model for GRBs in Section 5, with its luminous GeV-TeV radiation from the SSC process. Calculations from the ESM in a uniform surrounding medium establish quantitative predictions for the MeV, GeV, and TeV behavior of GRBs in Section 6 -- for one parameter set.  Finally, in Section 7 we mention possible high energy gamma-ray signatures of ultrahigh energy hadrons accelerated by GRB blast waves. We summarize in Section 8.
 
\section{Gamma Rays from Gamma Ray Bursts} 
 
\begin{figure}[b] 
\centerline{\epsfxsize=0.5\textwidth\epsfbox{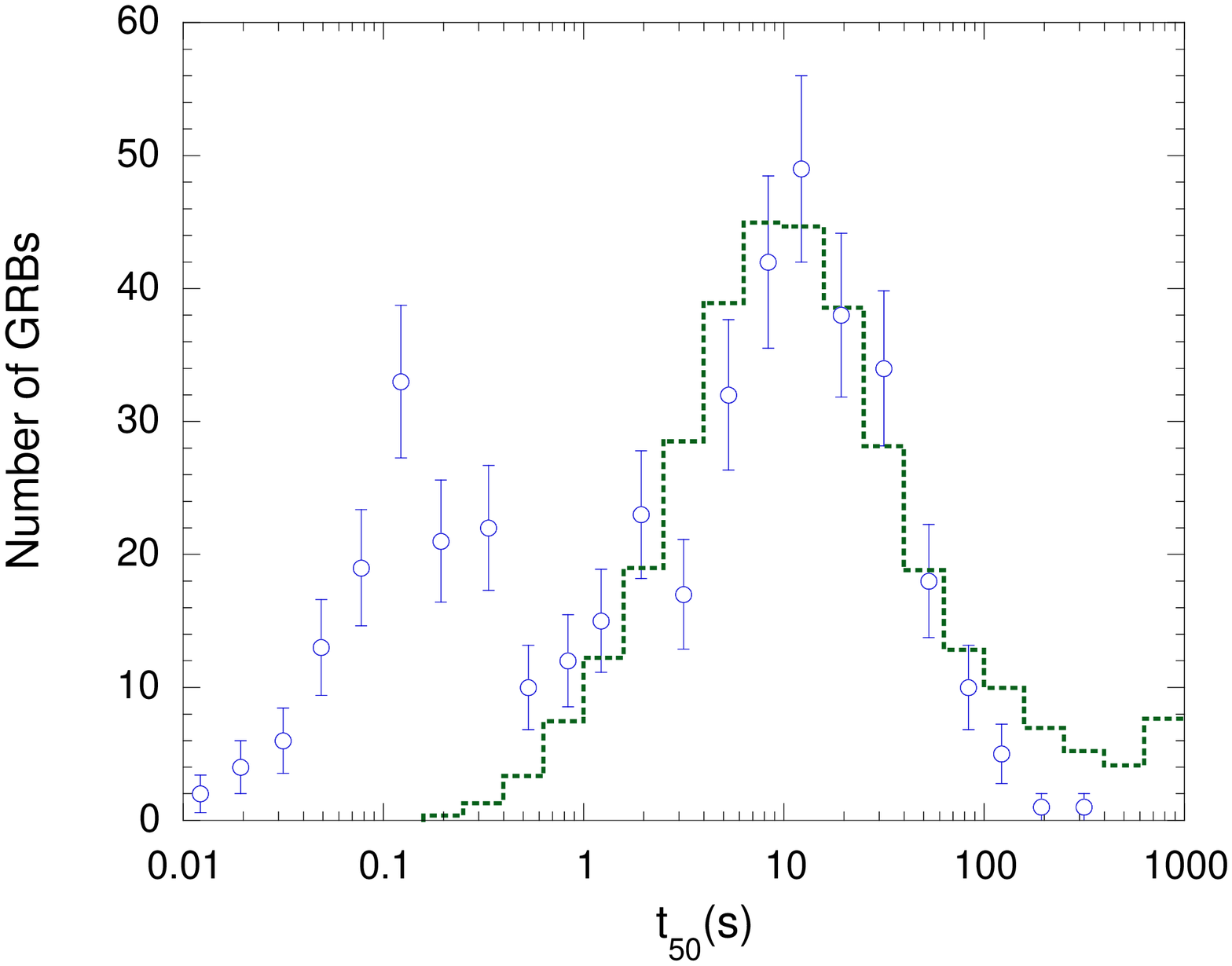} 
            \epsfxsize=0.5\textwidth\epsfbox{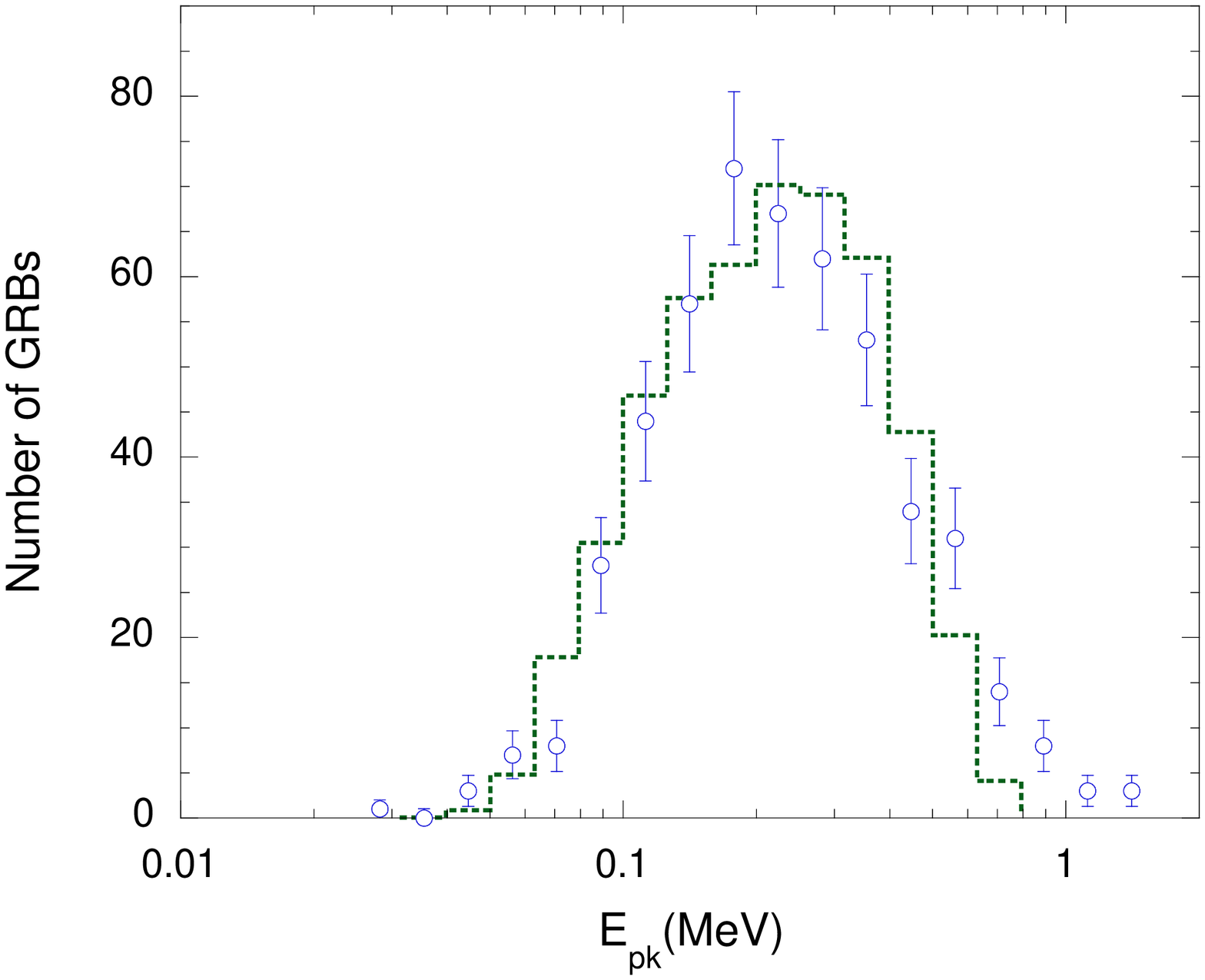}} 
\vskip-1.5in 
\caption[] {\baselineskip2pt 
Data points give the $t_{50}$ duration (left) and $E_{\rm pk}$ (right) distributions of GRBs measured with BATSE \cite{mea96,mallozziea98}. Dotted histograms give model fits from the external shock model \cite{bd00}.} 
\end{figure} 
 
The largest and most complete GRB data set has been obtained with the BATSE detector on the {\it Compton Observatory} \cite{mea96,mea98}. In its normal mode of operation, BATSE triggers when the 50-300 keV count rate in two detectors exceeds 5.5$\sigma$ over background on time scales of 64, 256, and 1024 ms.  The background is obtained from a commandable time interval, usually set at $\approx 17$ seconds. Fig.\ 2 shows the $t_{50}$ duration \cite{mea96} and $E_{\rm pk}$ \cite{mallozziea98} distributions measured with BATSE.  The $t_{50}$ duration is the time interval over which the integrated counts range from 25\% to 75\% of the total counts over background.  The value of $E_{\rm pk}$ is the photon energy of the peak of the time-averaged $\nu F_\nu$ GRB spectral energy distribution.  As can be seen, the duration distribution shows two distinct components \cite{kea93}; furthermore, there is a clear correlation for shorter GRBs to have harder spectra.  The range of $E_{\rm pk}$ is quite narrowly distributed in a range centered at $\sim 200$ keV, which is right in the middle of the triggering range of BATSE.  According to the ESM \cite{bd00}, $E_{\rm pk}$ is primarily determined by the baryon loading, and the $E_{\rm pk}$ distribution is a consequence of the triggering properties of BATSE convolved with the flux behavior of GRB blast waves with different total energies and baryon-loading factors which explode in surroundings with a range of densities.
 
The generic spectral form of GRB emission in the BATSE energy range is  
\begin{equation} 
{dN\over dE} \propto 
 \cases{ E^{-\alpha_{\rm ph}}\;, & for $E < E_{\rm pk}$ \cr 
	 E^{-\beta_{\rm ph}}\; & for $E > E_{\rm pk}$\cr	}\;  
\label{spct} 
\end{equation} 
where, typically, $\alpha_{\rm ph} \cong 1$ and $\beta_{\rm ph} = 2$-2.5. The {\rm Solar Maximum Mission} satellite revealed that $\gtrsim 1$ MeV emission was a common property of GRBs \cite{mea85}, thus establishing that the radiation has a nonthermal origin. COMPTEL has detected over 30 GRBs at $E > 0.75$ MeV \cite{connorsea98}. The spark chamber on EGRET detected $E\gtrsim 30$ MeV photons from 7 GRBs \cite{dcs98}. These GRBs are invariably among the brightest BATSE bursts. The average spectral index of four EGRET GRBs, consisting of 45 photons with energies $> 30$ MeV, is $\beta_{\rm ph} = 1.95\pm 0.25$ \cite{dcs98}, consistent with this emission being an extension of the spectrum near $E_{\rm pk}$ observed with BATSE.  EGRET's TASC, which measures $\sim 1$-200 MeV spectra and serves as a calorimeter to measure total photon energy for EGRET, has detected at least 16 GRBs \cite{cds98}.
 
\begin{figure} 
\centerline{\epsfxsize=1.1\textwidth\epsfbox{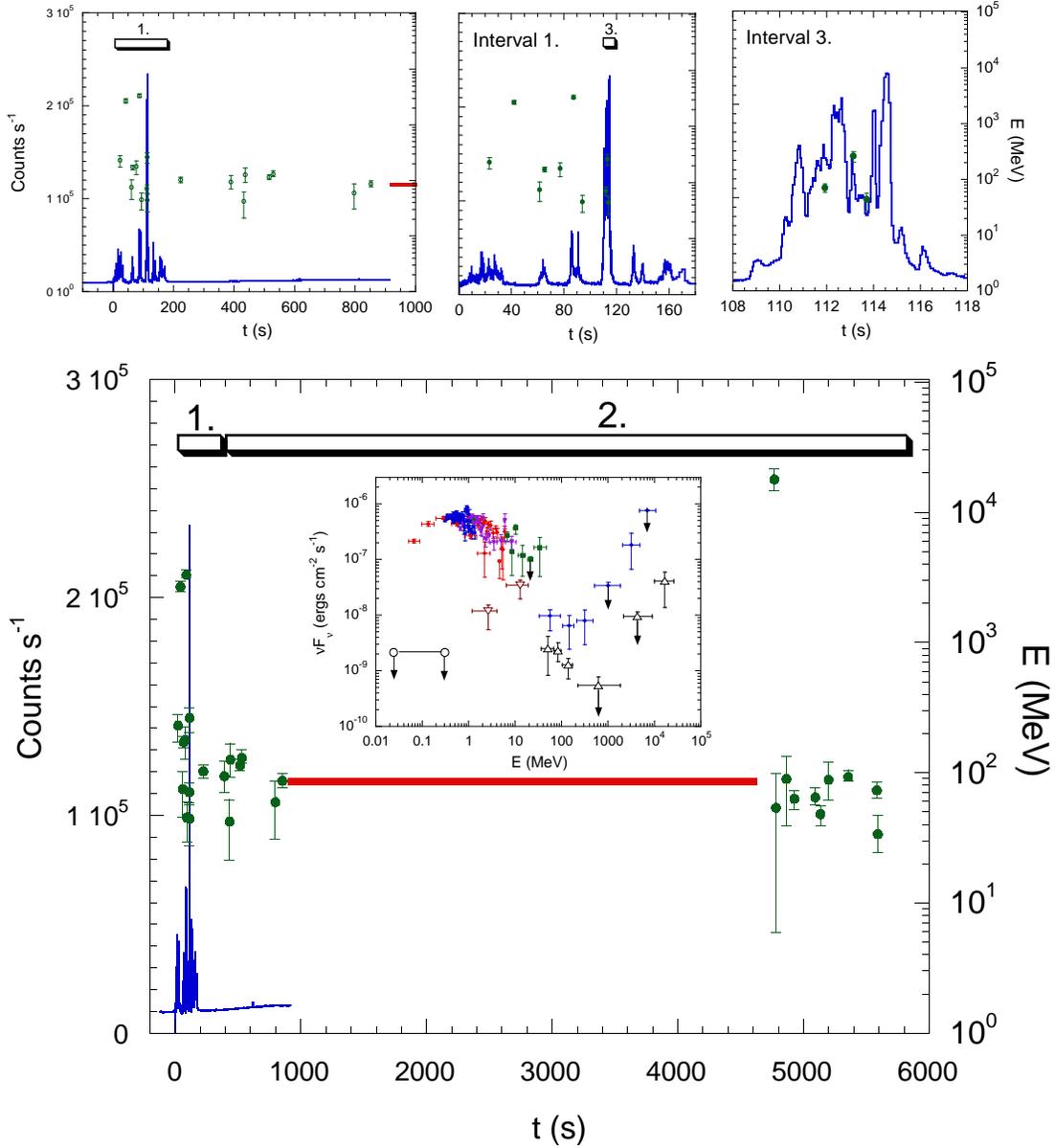}} 
\vskip-1.0in 
\caption[] {\baselineskip2pt 
Central figure: Times and energies of EGRET-detected photons and BATSE light curve of GRB 940217 (BATSE trigger \# 2831)\cite{hea94}. Inset shows composite $\nu F_\nu$ spectra during interval 1 and interval 2 (large symbols), naively obtained by multiplying photon spectrum by $E^2$. The BATSE light curve is summed 16 channel MER data (excluding channel 12 for which only the first 8 seconds of data exist). Successive blow-ups of the BATSE light curve are shown in the top panels.  Note that deadtime effects from background vetos could have reduced EGRET's efficiency for detecting gamma-rays. } 
\end{figure} 
 
Fig.\ 3 shows the light curve and spectra of the famous burst GRB 940217, which displayed an Earth-occulted $\sim 100$ MeV tail that lasted for $\sim 95$ minutes, two $\sim 3$ GeV photons during the BATSE burst, and an 18 GeV photon 90 minutes later\cite{hea94}. The Interval 1 and 2 $\nu F_\nu$ spectra are shown in the inset, and the three EGRET photons detected during the brief interval 4 are shown in the upper right panel. 
 
\section{Unsolved GRB Problems} 
 
If, as generally reasoned, GRB emissions originate from a fireball that ejects either a single blast wave into inhomogeneous surroundings or expels a long-lasting relativistic wind, then a central problem in GRB studies is to understand the nature of the central engine and how it powers the energy  released into the blast wave.  The favored, but by no means proven scenario is that GRBs are powered by the core collapse of a massive star to a black hole. The short events (i.e., $t_{50} \lesssim 1$ s in Fig.\ 2a) may have a separate origin, for example, through compact object coalescence.  A massive star origin for GRBs is in accord with the vigorous star formation implied by the blue galaxy hosts, the evidence for large quantities of gas and dust in GRB environs, and the coincidence of GRB directions with the disk and central regions of host galaxies.
 
The degree of GRB blast-wave collimation remains a crucial unknown. Neither compact object coalescence scenarios nor collapsar/hypernova models invoking neutrino annihilation or poorly quantified mhd processes make sufficient fireball energy to account for the largest measured GRB energies without invoking opening half-angles $\psi \lesssim 10^\circ$ (e.g.,\cite{jea99,pwf99}). Easing the energy requirements is a great boon to these and other models. 
 
A third open question is whether the prompt GRB emission results from collisions between a succession of shells ejected from the GRB engine \cite{rm94,kps97} or is instead due to an ESM where a single impulsive relativistic blast wave interacts with inhomogeneities in the external medium \cite{mr93,dm99}. The answer to this problem characterizes the accretion/collapse and coalescence activity taking place near GRB engines. 
 
\section{$\gg$ Transparency Arguments} 
 
Gamma-ray observations set important constraints on the location and speed of the blast wave shell through the requirement that the emission region be optically thin to $\gg$ pair production attenuation \cite{kp91,feh92,bh97}.  We estimate the optical depth $\tau_{\gamma\gamma}(\ep)$ to pair production at dimensionless photon energy $\ep = h\nu^\prime/m_ec^2$. Primed quantities refer to the comoving blast wave frame and unprimed quantities refer to the observer frame. We have 
\begin{equation} 
\tau_{\gamma\gamma}(\ep) \cong \bigl[ {\ep L(\ep )\over \ep m_ec^2 }\cdot {\Delta R^\prime\over c}\cdot {1\over 4\pi R^2 \Delta R^\prime}\bigr] \cdot\sigma_{\gamma\gamma}(\ep ) \cdot \Delta R^\prime\ \, . 
\end{equation} 
The blast wave shell, with comoving width $\Delta R^\prime$, is at distance $R$ from the explosion site when it radiates the photons that are measured with gamma-ray detectors at energies $\e \cong \Gamma\ep /(1+z)$.  The total power gets boosted and redshifted by two factors of energy and time for a spherically expanding blast wave; thus $\epsilon L(\epsilon) = \Gamma^2 \ep L(\ep)/(1+z)^2$ and $\e \cong \Gamma\ep/(1+z)$. By definition, $4\pi d_L^2 \e S(\e ) = \e L(\e )$, where $d_L$ is the luminosity distance and $S(\e )$ is the spectral flux (ergs cm$^{-2}$ s$^{-1} \e^{-1}$). 
 
The $\gg$ cross section peaks near threshold and reaches a value of $\sigma_\gg (\epsilon'\sim 1)\approx \sigma_{\rm T}$. An estimation of merely the $\gg$ optical depth of near-threshold-energy photons in the blast wave frame -- which are detected with  $\e \cong \Gamma/(1+z)$ -- gives  
\begin{equation} 
\tau_\gg [\e = \Gamma /(1+z)] \simeq ({ 1+z \over  \Gamma })^2 \;\;{ d_L^2 S_0 \e^{1-\alpha}\Delta R^\prime \sigma_{\rm T} \over R^2 m_ec^3}\,. 
\end{equation} 
Here we parameterize the observed high-energy photon spectrum $\nu F_\nu = \epsilon S(\e) = S_0 \e^{-\alpha}$. Requiring $\tau_\gg < 1$ and invoking the relation $\Delta R^\prime = fR/\Gamma$, where $f\sim 1$ for an adiabatic blast wave \cite{mlr93,bm76}, we place limits on the Lorentz factor $\Gamma$ and the location $R$ of the site where high energy radiation is produced.  Suppose that a power-law spectrum of $\gamma$ rays extending to energy $\epsilon_{\rm max}$ is measured. Then either $\Gamma \gtrsim (1+z)\epsilon_{\rm max}$, or 
\begin{equation} 
R \gtrsim {d_L^2\over (1+z)}\; {{S_0 \sigma_{\rm T} f \over \epsilon_{\rm max}^{\alpha+2} m_ec^3 }}\;= 2.7\times 10^{21} {S_{-6} d_{28}^2 f\over (1+z) \e_{\rm max}^{\alpha + 2}}\; {\rm cm}\, , 
\end{equation} 
where $S_{-6}= S_0/10^{-6}$ ergs cm$^{-2}$ s$^{-1}$ and $d_{28} = d_L/10^{28}$ cm.   
 
For the specific case of GRB 990123 shown in Fig.\ 1, $\epsilon S(\epsilon) = 6.7\times 10^{-6} \e^{-1.1}$ ergs cm$^{-2}$ s$^{-1}$, so that $S_{-6} = 6.7$ and $\alpha = 2.1$. Furthermore, $d_{28} = 3.1 $ (see Table 1).  The BATSE and COMPTEL observations of 4-8 MeV photons already imply that either $\Gamma > 20$-40 or $R\gtrsim 5\times 10^{18}f$ cm.  If 100 MeV photons had been observed coincident with this GRB (unfortunately, EGRET's spark chamber did not observe this GRB as it was too far off axis), then we could draw the conclusion that either $\Gamma \gtrsim 500$ or $R\gtrsim 3\times 10^{13}f$ cm. This can restrict some forms of the internal shock model \cite{wb97}, with implications for neutrino production by GRBs. 
 
Application to GRB 940217 provides looser constraints on $R$ and $\Gamma$ because we do not know its redshift, again highlighting the importance of redshift measurements. The $\gamma\gamma$ transparency constraints can be strengthened when one considers pair-producing interactions between high energy $\gamma$ rays and lower energy photons\cite{baring99}.  The use of gamma-ray astronomy to infer properties of the expanding outflow will be well utilized by future AGILE and GLAST observations in the 100 MeV - GeV range, and also potentially from $\sim 0.1$-1 TeV emission observed with ground-based air or water \^Cerenkov telescopes. 
 
\section{External Shock Model for GRBs} 
 
A minimum of nine parameters enter into a blast-wave model calculation for GRBs in the ESM (for details and references on the next 2 sections, see \cite{cd99,dcm99}). These can be grouped according to whether they are (i) intrinsic parameters associated with the properties of the central engine, (ii) environmental parameters that characterize the surrounding medium, or (iii) microscopic parameters that define the reinjection of swept-up hadron power into the nonthermal leptons in the blast wave.  
 
The three intrinsic parameters are the directional energy $\partial E_0/\partial \Omega \rightarrow  E_0/(4\pi) = 10^{54}E_{54}$ ergs/($4\pi$ sr) released by the central engine, the initial Lorenz factor $\Gamma_0$ of the blast wave, and the opening half-angle $\psi$ of the collimated outflow.  We take $E_{54} = 1$ and consider either uncollimated or collimated outflows with $\psi = 10^\circ$. This opening angle relaxes the energy requirements by a factor of $\sim 130$ for a one-sided jet. (Two additional complications, not dealt with here, are angular gradients in outflows and lateral spreading of the blast wave.)
 
The initial blast wave Lorentz factor $\Gamma_0$ is closely related to the baryon loading of the fireball, because the optically thick fireball expands until most of its initial energy $E_0$ has been transferred to the kinetic energy of the outflowing baryons $\Gamma_0 M_{\rm b} c^2$, where $M_{\rm b}$ is the mass of the baryons. As the blast wave sweeps up and captures material from the surrounding environment, it decelerates and becomes energized by the addition of nonthermal particles with Lorentz factors $\Gamma$ in the comoving blast-wave frame.  The circumburster environment is likely to be highly structured in all cases, but especially if the progenitor of a GRB is a massive star located in a star forming region where stellar winds could introduce inhomogeneities. Nevertheless, the surrounding density distribution is usually parameterized by the function $n(x) = n_0 x_{\rm dec}^{-\eta}$. We take $n_0 = 100$ cm$^{-3}$ and $\eta = 0$ as standard values, though $\eta = 2$ would be more appropriate for a wind. In the ESM, the measured durations of GRBs are comparable with the deceleration time scale $t_{\rm dec} = [3(\partial E_0/\partial \Omega)/m_pc^2 n_0]^{1/3}/(c\Gamma_0^{8/3})$.  
 
The microscopic parameters include the fraction $\epsilon_e$ of nonthermal swept-up proton kinetic energy transferred to nonthermal electrons, the injection index $p$ of the electrons, and the maximum electron energy parameter $\epsilon_{\rm max}$, given through the kinematic limit $\gamma_{\rm max}= 4\times 10^7 \epsilon_{\rm max}/\sqrt{B ({\rm G})}$. The comoving magnetic field strength $B$ is set by an equipartition argument. The value of the magnetic equipartition parameter $\epsilon_B$ is defined by $B^2/8\pi = 4\epsilon_B (\Gamma^2-\Gamma)m_pc^2 n(x)$. We let $\epsilon_e = 0.5$, $\epsilon_{\rm max} = 1$, $p=2.5$, and $\epsilon_B = 10^{-4}$, and furthermore assume that the microscopic parameters are time-independent.  The low value of $\epsilon_B$ is required \cite{cd99} to reproduce the generic eq.(1) spectrum. 
 
\section{Model Spectra and Light Curves} 
 
The numerical simulation model \cite{dcm99} treats synchrotron, synchrotron self-Compton (SSC), synchrotron self-absorption and adiabatic loss processes, and follows blast-wave evolution self-consistently. The photons are attenuated by $\gg$ absorption, but pair reinjection is not followed. Fig.\ 4 shows temporally evolving spectra for the standard uncollimated parameter set. The $\gg$ process degrades only $\gtrsim$ 0.1 TeV photons for the results shown here. Thus the internal attenuation of high-energy gamma rays in the ESM is not too severe and the SSC component is bright enough that TeV radiation is produced at a comparable $\nu F_\nu$ level as the synchrotron radiation. Dirty fireballs produce a larger relative $\nu F_\nu$ flux in the SSC component than in the synchrotron component. We \cite{dcm99} propose that the TeV radiation detected by Milagro from GRB 970417a and reported at this meeting \cite{mea99} is the SSC emission from a nearby $z\lesssim 0.1$ GRB. 
 
Fits to Fig.\ 2 data \cite{bd00}, taking into account BATSE triggering properties and the strong biases against detecting dirty fireballs with $\Gamma_0 \ll 300$, imply a very large population of undiscovered optical and X-ray transients with well-characterized properties.  The clean fireball population produces sub-second transients peaking at GeV-TeV energies \cite{dcb99} that GLAST, AGILE, or \^Cerenkov detectors could discover.  The distribution of baryon-loading parameters $\Gamma_0$ toward clean fireballs falls, however, below a power-law parameterization of the $\Gamma_0$ distribution \cite{bd00}, indicating that the space density of clean fireballs is less than that for fireballs producing detectable GRBs ($\Gamma_0 \sim 300$). Very clean fireballs ($\Gamma_0 \gtrsim 3000$) could produce TeV bursts of radiation. These can be distinguished from Hawking radiation and annihilating dark matter particles by their spectrum and afterglow.

\begin{figure}[t] 
\centerline{\epsfxsize=0.35\textwidth\epsfbox{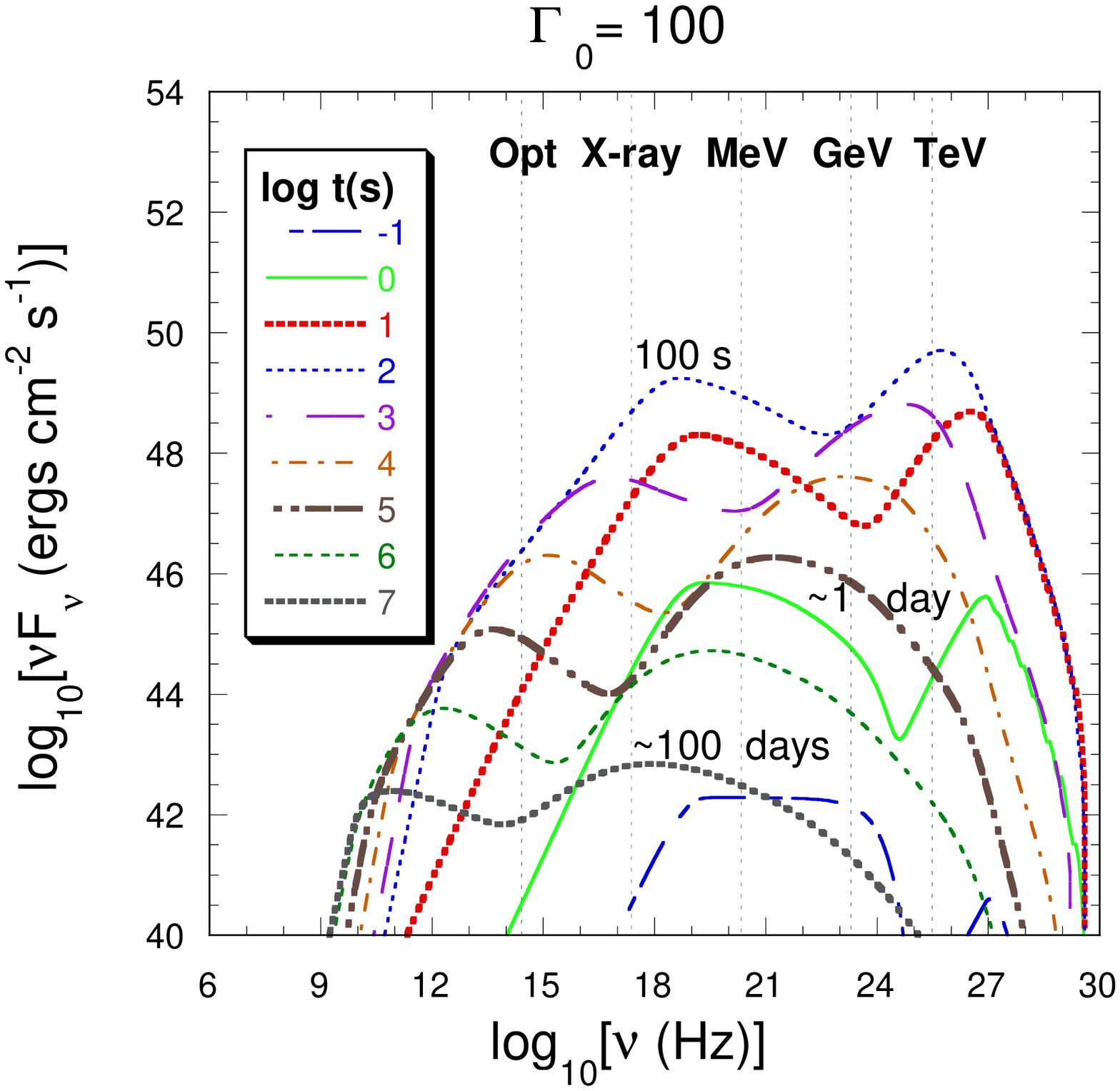} 
		\epsfxsize=0.35\textwidth\epsfbox{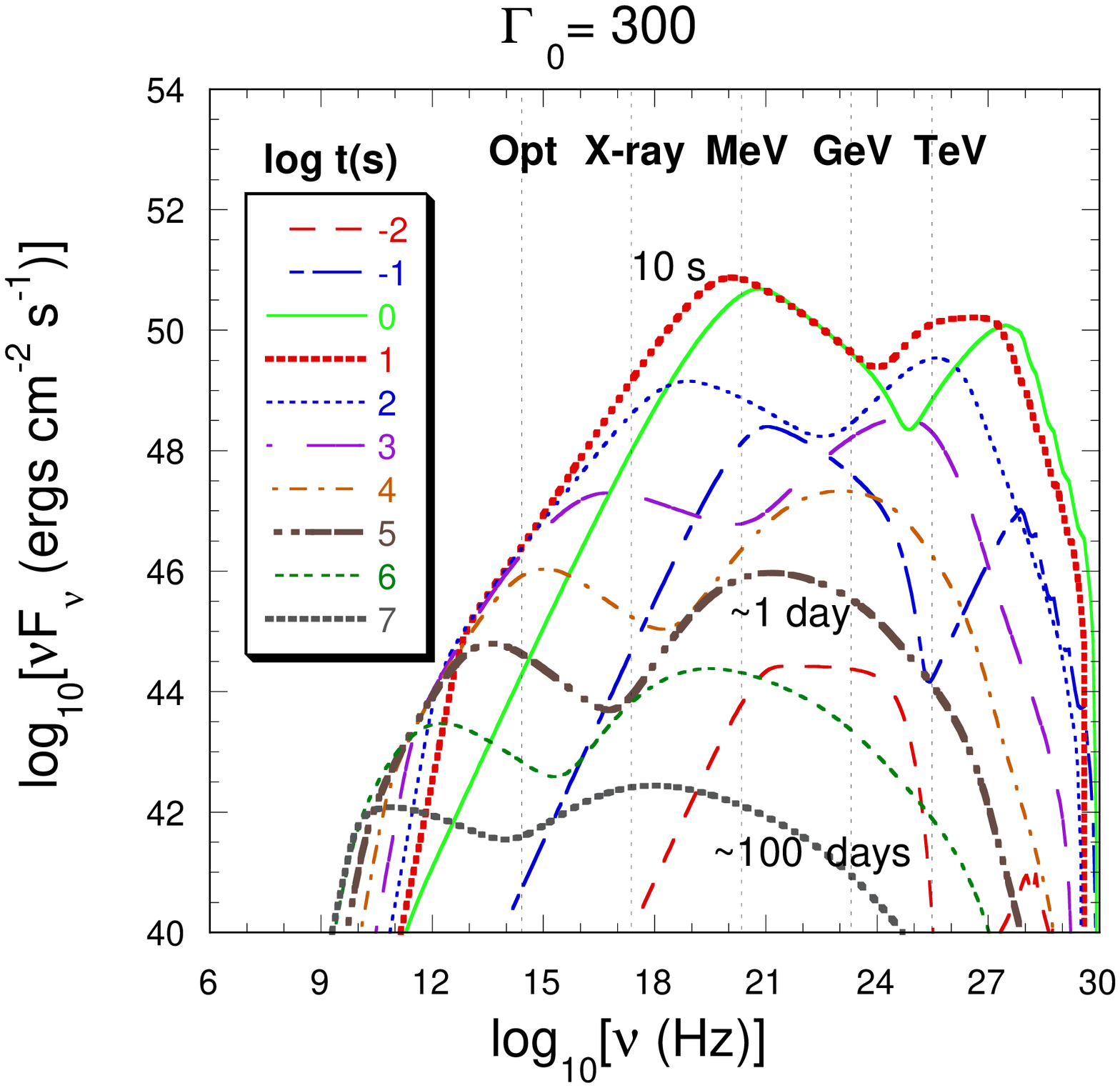} 
            \epsfxsize=0.35\textwidth\epsfbox{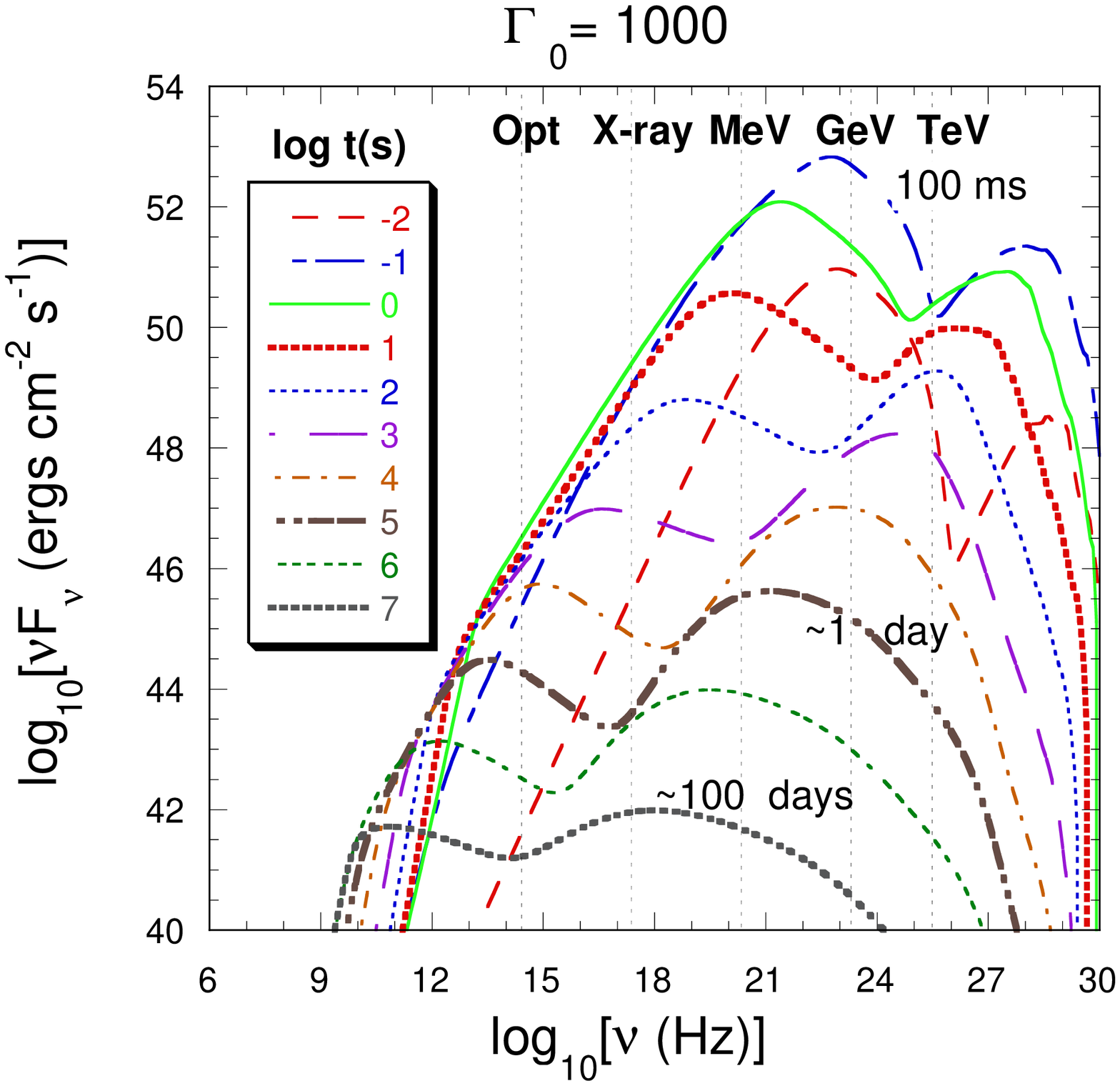}} 
\vskip-0.5in 
\caption[] {Calculations of SEDs from uncollimated GRB blast waves that are energized, decelerate and radiate by capturing material from a uniform surrounding medium with \S V parameters. Only the initial Lorentz/baryon-loading factor $\Gamma_0$ differs between the three calculations. The duration decreases and the $\nu F_\nu$ flux and $E_{\rm pk}$ values increase with increasing $\Gamma_0$. } 
\end{figure} 
 
\begin{figure}[b] 
\centerline{\epsfxsize=0.5\textwidth\epsfbox{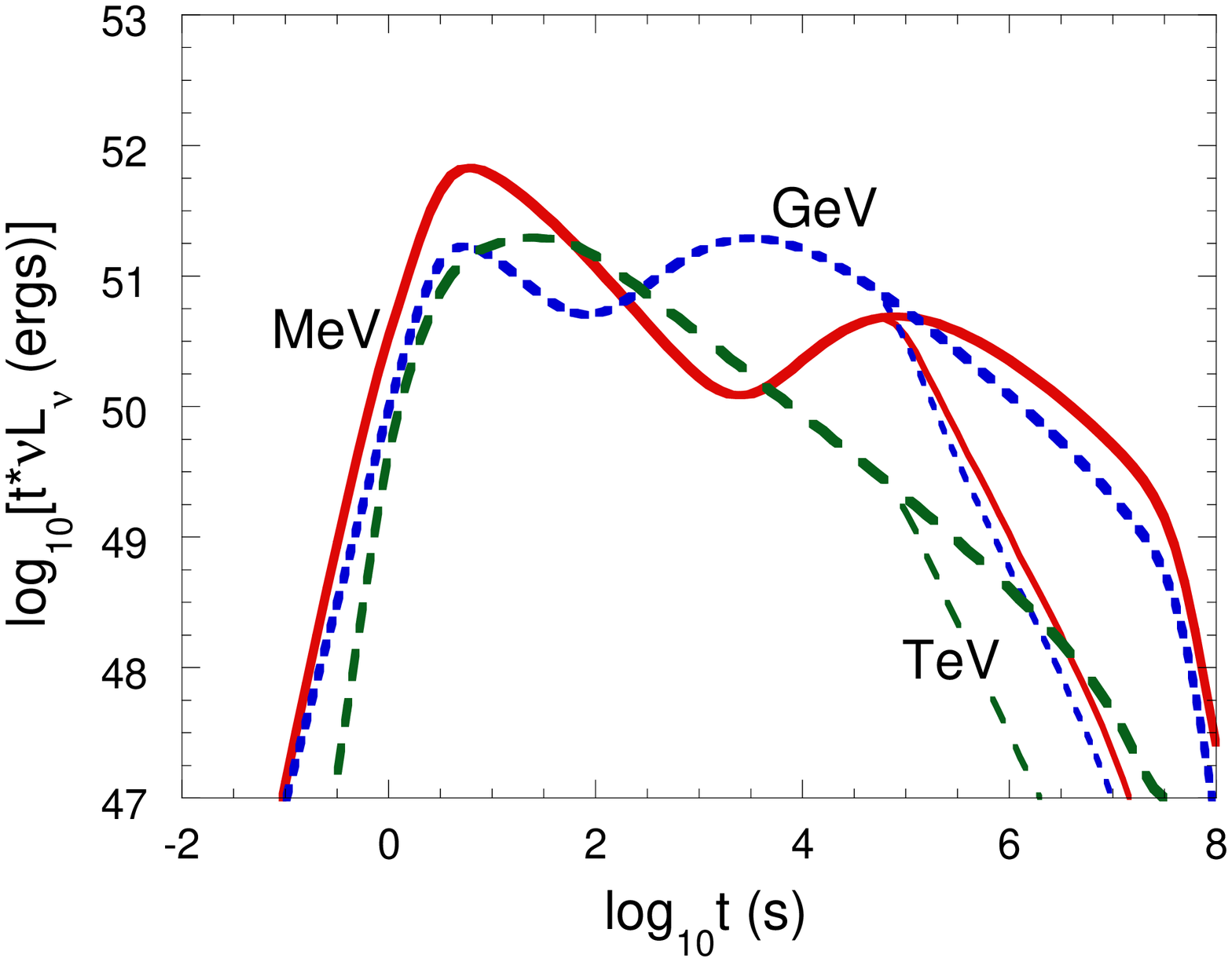} 
		\epsfxsize=0.5\textwidth\epsfbox{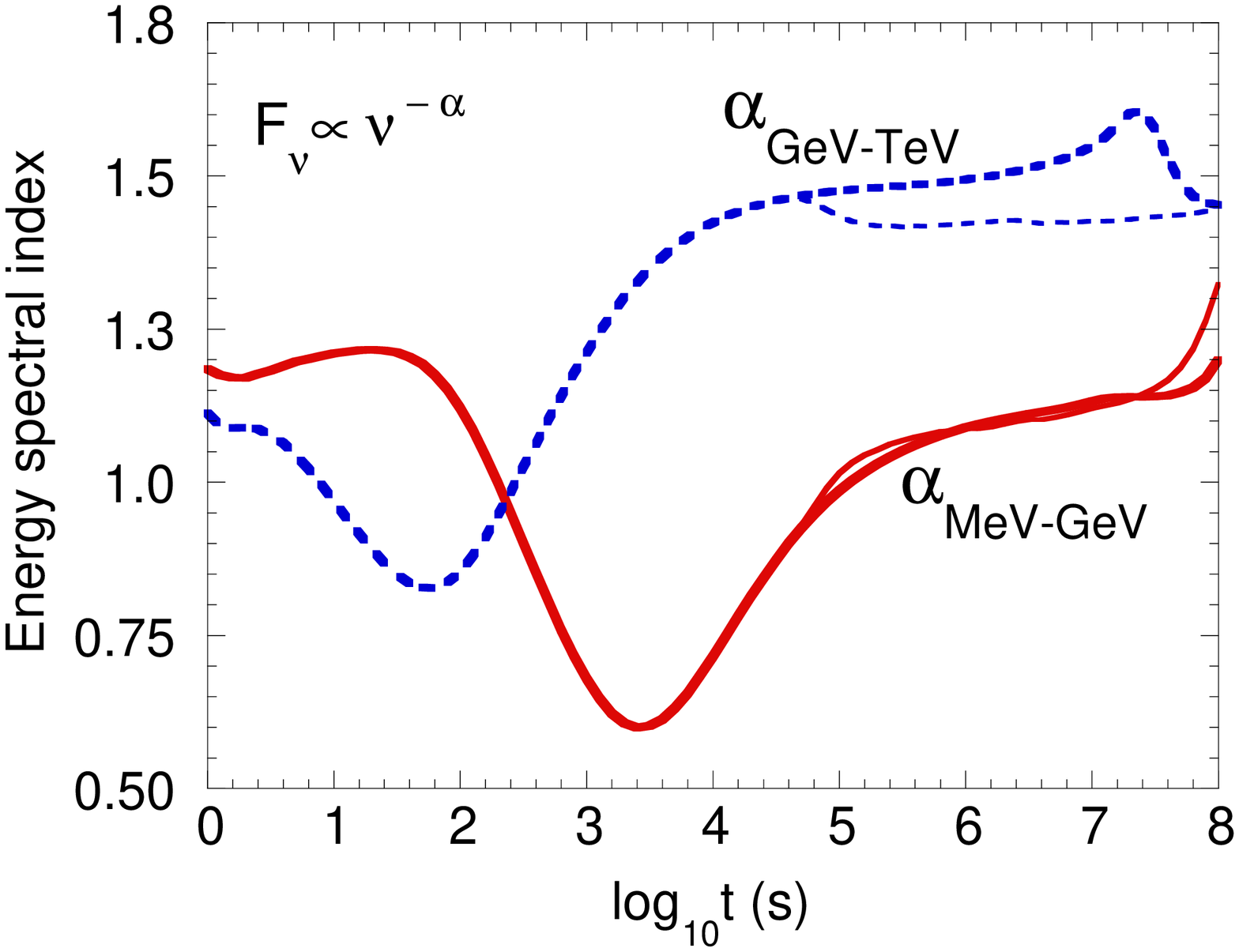}} 
\vskip-1.5in 
\caption[] {(left) Product of $\nu L_\nu$ flux and observing time $t$ for the MeV, GeV, and TeV light curves using $\Gamma_0 = 300$ case in Fig.\ 4.  (right) Temporal variation of the broadband MeV-GeV and GeV-TeV energy spectral indices. In both panels, thick curves are for uncollimated outflows and thin curves are for beamed outflows with $\psi = 10^\circ$ and an observer along the symmetry axis of the jet.} 
\end{figure} 
 
The MeV, GeV, and TeV $\nu L_\nu$ light curves shown in Fig.\ 5 \cite{dcm99} are multiplied by time to show where most counts will be detected in logarithmic intervals of time. Synchrotron radiation forms the early MeV and GeV peaks -- this is the GRB itself.  The SSC component forms the early TeV peak. The later peaks at MeV and GeV energies are due to the SSC component becoming increasingly dominant in these wavebands as the blast wave decelerates. The second maximum at GeV energies occurs at $\approx 5000$ s, comparable to the duration of the extended emission observed from GRB 940217.  Because the relative fluxes of prompt and delayed emission are greater at GeV energies than at MeV energies for these parameters, it is more probable that delayed GeV emission rather than MeV emission would be detected from a GRB, as in fact was observed with EGRET from GRB 940217. However, the particular calculation shown here corresponds to a uniform surrounding, whereas the behavior of the light curve shown in Fig.\ 3 could be explained with the ESM only in terms of a highly structured medium. We emphasize that the full range of possible behaviors for spectral and temporal evolution, of which Fig.\ 5 represents only one possibility, has hardly been explored. 
 
GLAST, with its larger effective area and field-of-view, should be able to monitor the evolution of the SSC spectral feature due to blast wave deceleration from many bright GRBs. Broadband MeV-GeV and GeV-TeV spectral indices due to blast-wave deceleration are plotted in the right panel in Fig.\ 5 for the standard parameter set studied here. During the early phase, the MeV-GeV photon spectral index corresponds to a soft cooled synchrotron spectrum, here with a value $\beta_{\rm ph} \sim 2.25$ for the $p=2.5$ injection electron spectrum, in accord with measurements of $> 30$ MeV EGRET spectra \cite{cds98}. The GeV-TeV index is much harder because this waveband primarily samples the harder SSC component. After the prompt phase, the GeV-TeV index softens to a spectrum that is even softer than the cooled synchrotron spectrum due to effects of $\gg$ attenuation, and the MeV-GeV index hardens as the SSC radiation sweeps into this waveband.  The MeV-GeV index approaches the cooled synchrotron limit at later times.  Spectral hardening in the MeV-GeV band in the early afterglow phase due to the deceleration of the blast wave as it interacts with a smooth external medium constitutes generic behavior of the ESM which can be tested with GLAST, and can be confronted by observations of GeV and TeV detectors of bright GRBs with smooth MeV light curves that signify a GRB source within a uniform surrounding. 
 
\section{Hadrons in GRB Blast Waves} 
 
The nonthermal energy carried into GRB blast waves by hadrons is larger by a factor $\sim m_p/m_e$ than the energy carried by leptons, so hadronic effects can hardly be negligible. The physics of transferring energy from hadrons to leptons is just one of the many open questions in this field.  An important related question is whether GRB blast waves accelerate ultra-high energy cosmic rays (UHECRs). The validity of this idea \cite{v95,w95,mu96} was argued by comparing the energy densities of  UHECRs with the globally averaged injection of energy by GRBs into a volume no greater than, for $\gtrsim 10^{20}$ eV UHECRs, the Zatsepin-Kuzmin-Greisen radius outside which UHECRs are degraded by photomeson production on cosmic microwave background photons.  Photomeson neutrino production at $\gtrsim 10^{14}$ eV \cite{wb97}, and GeV $\gamma$-ray production from proton synchrotron radiation \cite{v97a,bd98} are both potentially observable signatures of UHECR acceleration by GRBs. 
 
The UHECRs are claimed to be accelerated either through a first-order shock \cite{v95} or second-order \cite{w95} stochastic Fermi process. The shock Fermi mechanism fails for collapsar models of GRBs \cite{ga99} because only the first shock cycle produces a $\Gamma^2$ energy gain. Subsequent cycles give energy increases of only factors-of-2, because the shock catches up to the particle before it can complete more than a small fraction of its cycle.   
 
The simplest approach is to assume \cite{ps99,pohl99} that no acceleration follows the capture of particles into the blast wave; of course, no UHECRs are then produced. Such a process could produce a low-level flux of $\gamma$ rays at $E \sim 0.1 \Gamma/(1+z)$ GeV and radio/optical synchrotron radiation from the process $p+p\rightarrow \pi + X \rightarrow \gamma$, e$^\pm +X $, where the low-energy protons are the thermal baryon-load material. This approach appears too inefficient, however, to describe flaring events. Magnetic turbulence injected by charged dust during the capture and isotropization process could, though gyroresonant processes, accelerate protons to ultra-high energies \cite{sd99}, as could the turbulence generated when the blast wave encounters inhomogeneities in the circumburster medium. The shock front will likely be Rayleigh-Taylor unstable, and this will also generate turbulence in the blast wave.
 
The population of GRBs with redshifts now permits a more quantitative estimate of the rate density of GRB sources. Stecker \cite{stecker99} argues that if GRBs follow the star-formation rate history of the universe, then a much smaller energy injection rate of UHECRs into the local universe occurs, so that GRBs cannot be the source of the UHECRs. This argument does not take into account the predicted but so-far undetected dirty fireball population, which can introduce a 2-3 orders-of-magnitude increase in the source density of GRBs \cite{bd00} and therefore UHECRs. It has also been argued \cite{totani98} that if energy is transferred very inefficiently from hadrons to electrons, then the total hadron energy in GRBs is $\sim m_p/m_e$ greater than implied by the gamma-ray measurements. Hence $\partial E/\partial \Omega\rightarrow 10^{56}$ ergs sr$^{-1}$. We resist this proposal because it multiplies difficulties in understanding the energetics of GRB sources.  
 
Slow decay of GeV-TeV radiation from proton synchrotron radiation provides evidence in favor of hadrons in GRBs. Protons are much less radiative than leptons unless they are far more energetic; thus hadrons are more likely to be weakly cooled. When protons are injected with number index $ s = 2$, the uncooled GeV proton synchrotron flux decays in the adiabatic regime with temporal index $\chi = 3/4$ (flux $\phi \propto t^{-\chi}$). In comparison, the optical and X-ray synchrotron radiation decays as $\chi = 1$ for strongly cooling electrons that are likewise injected with $s = 2$.  The temporal decay from slowly cooling hadrons is thus slower than for strongly cooling leptons. Consequently GeV proton synchrotron radiation should decay more slowly than lepton synchrotron and SSC radiations.  Before more concrete conclusions can be drawn, however, further studies are needed to distinguish between the behavior of the hadronic and SSC emissions, and to treat diffusive acceleration, cascade processes and UHECR escape from the GRB blast wave. 
 
\section{Summary} 
 
Gamma-ray transparency arguments push one irresistibly toward a relativistic blast-wave model of GRBs. The standard fireball/blast wave model implies strong GeV/TeV radiation from the SSC process \cite{dcm99}. Using parameters optimized to fit prompt hard X-ray and soft $\gamma$-ray emission from GRBs, our calculations show nearly coincident MeV/TeV light curves and extended GeV light curves due to the dominance of the SSC component at GeV energies in the early afterglow phase.  In the framework of the ESM, GeV and TeV observations chart the evolution of the SSC component and hence the evolution and changes of the blast wave. Calculations of the MeV, GeV and TeV light curves were made for a standard parameter set, showing that the GeV band displays a soft-to-hard-to-soft evolution as the SSC component sweeps through this waveband.  This spectral prediction applies to blast waves which decelerate in a uniform medium as evidenced by smooth GRB light curves; circumburster medium structure introduces many possible variations to the light curves and spectral behaviors not yet explored (compare Figure 3).   
 
The possible existence of a class of very clean fireballs that produce $\lesssim 100$ ms flashes of GeV and TeV radiation is a straightforward prediction of the blast wave model. The related prediction of a large class of dirty fireballs finds good company with the hypothesis that GRB blast waves accelerate energetic hadrons, because the dirty fireballs provide a much more numerous source population with which to provide the energy of the UHECRs observed locally. Hadronic acceleration might reveal itself through the slow decay of GeV-TeV proton synchrotron radiation, but better studies are needed for quantitative predictions.  
 
\acknowledgments 
 
CD thanks Anthony Crider for help in preparing Fig.\ 3 and for discussions about the BATSE data. He also thanks conference organizers Brenda Dingus and Mike Salamon for an exciting meeting.  We acknowledge collaboration and useful discussions with M. B\"ottcher.

\end{document}